\documentclass[english]{article}
\usepackage[table]{xcolor}
\usepackage{amsmath}
\usepackage{amsfonts}
\usepackage{amssymb}
\usepackage{mathtools}
\usepackage{array}
\usepackage{longtable}
\usepackage{soul} 
\usepackage[hidelinks]{hyperref}
\usepackage{textcomp}
\usepackage{float}
\usepackage[margin=1in]{geometry}
\usepackage[utf8]{inputenc}
\usepackage{wrapfig}
\usepackage[big]{titlesec}
\usepackage{subcaption}
\usepackage{caption}
\usepackage{bm}
\usepackage{mathrsfs}
\usepackage[en-US]{datetime2}
\usepackage{svg}
\usepackage{cleveref}
\usepackage{booktabs}
\DTMlangsetup[en-US]{showdayofmonth=false}
\usepackage{setspace}
\usepackage{upgreek}
\usepackage{tabularx}
\usepackage{makecell}
\usepackage{multirow}
\usepackage{multicol}
\usepackage[sort&compress]{natbib}
\usepackage{authblk}
\usepackage{tikz}
\usetikzlibrary{positioning} 
\usetikzlibrary{matrix}
\usetikzlibrary{patterns}
\usetikzlibrary{fit}
\usetikzlibrary{shapes, arrows.meta}

\usepackage[sort&compress]{natbib}

\newcolumntype{Y}{>{\centering\arraybackslash}X}

\newcommand{\bmat}{\begin{bmatrix}}
\newcommand{\emat}{\end{bmatrix}}
\newcommand{\vectA}[1]{\text{\textbf{#1}}}
\newcommand{\vectG}[1]{\bm{#1}}
\newcommand{\up}[1]{\text{#1}}
\newcommand{\vbeta}{\bm{\beta}}
\newcommand{\vPsi}{\bm{\Psi}}
\newcommand{\vpsi}{\bm{\nu}}

\newcommand{\paren}[1]{\left( #1 \right)}
\newcommand{\bracket}[1]{\left[ #1 \right]}

\titlespacing{\section}{0pt}{3pt}{3pt}
\titlespacing{\subsection}{0pt}{3pt}{3pt}
\titlespacing{\subsubsection}{0pt}{3pt}{3pt}

\title{The Mixed Aggregate Preference Logit Model: A Machine Learning Approach to Modeling Unobserved Heterogeneity in Discrete Choice Analysis}

\author[1,2]{Connor R. Forsythe\thanks{E-mail: \texttt{cforsyth@andrew.cmu.edu}}}
\author[3]{Cristian Arteaga\thanks{E-mail: \texttt{arteagas@unlv.nevada.edu}}}
\author[4]{John P. Helveston\thanks{E-mail: \texttt{jph@gwu.edu}}}

\affil[1]{Department of Engineering and Public Policy, Carnegie Mellon University}
\affil[2]{Wilton E. Scott Institute for Energy Innovation}
\affil[3]{Department of Civil and Environmental Engineering and Construction, University of Nevada Las Vegas}
\affil[4]{Department of Engineering Management and Systems Engineering, The George Washington University}
\date{\today}

\begin{document}

\maketitle

\begin{abstract}
    This paper introduces the Mixed Aggregate Preference Logit (MAPL, pronounced ``maple'') model, a novel class of discrete choice models that leverages machine learning to model unobserved heterogeneity in discrete choice analysis. The traditional mixed logit model (also known as ``random parameters logit'') parameterizes preference heterogeneity through assumptions about feature-specific heterogeneity distributions. These parameters are also typically assumed to be linearly added in a random utility (or random regret) model. MAPL models relax these assumptions by instead directly relating model inputs to parameters of alternative-specific distributions of aggregate preference heterogeneity, with no feature-level assumptions required. MAPL models eliminate the need to make any assumption about the functional form of the latent decision model, freeing modelers from potential misspecification errors. In a simulation experiment, we demonstrate that a single MAPL model specification is capable of correctly modeling multiple different data-generating processes with different forms of utility and heterogeneity specifications. MAPL models advance machine-learning-based choice models by accounting for unobserved heterogeneity. Further, MAPL models can be leveraged by traditional choice modelers as a diagnostic tool for identifying utility and heterogeneity misspecification.
\end{abstract}

\vspace{1cm}

\begin{center}
    \textit{This work is currently under peer review. Comments and suggestions are welcome.}
\end{center}


\newpage

\section{Introduction}\label{sec:Intro}

Discrete choice models are leveraged in a wide range of research communities, including Transportation, Economics, Healthcare, Business, and others \citep{Train2009,Haghani2021}. The mixed logit model---alternatively known as the mixed multinomial, random parameters logit, or random coefficients logit model---is particularly important due to its capability, as proven by \citet[p.447]{McFadden2000}, to approximate choice probabilities from ``any discrete choice model derived from random utility maximization.'' However, a limitation of this theoretical work is that ``it provides no practical indication of how to choose parsimonious mixing families'' \citep[p.452]{McFadden2000}. This means modelers must make simplifying assumptions about how they believe heterogeneity should be distributed across a population for each feature in the model, including assuming independent mixing distributions, convenient mixing distributions (e.g., normal or log-normal), or preference homogeneity for a subset of product features. Furthermore, available software packages for estimating mixed logit models only support a limited number of heterogeneity distributions to choose from, e.g. \citep{Helveston2023, Arteaga2022}. Consequently, mixed logit's theoretical potential as a flexible approach for modeling unobserved preference heterogeneity often remains unrealized in practice, and many papers leveraging random utility-based mixed logit models make limiting assumptions \citep{Forsythe2023,Helveston2015, Philip2021, Guo2021,Kavalec1999, Revelt1998}.

This paper introduces the \textbf{M}ixed \textbf{A}ggregate \textbf{P}reference \textbf{L}ogit (MAPL, pronounced ``maple'') model, a novel class of discrete choice models that aims to relax many of the assumptions required in traditional mixed logit. MAPL models enable the ability to flexibly model various decision-making paradigms (e.g., utility maximization or regret minimization), heterogeneous consumer preferences, and governing functional forms via a new conception of unobserved heterogeneity. Whereas traditional discrete choice modeling frameworks like mixed logit parameterize preference heterogeneity through assumptions about feature-specific heterogeneity distributions, MAPL models directly relate model inputs to parameters of alternative-specific distributions of aggregate preference heterogeneity. In doing so, MAPL models also eliminate the need to specify a functional form for the latent decision model (e.g. a linear-in-parameter utility model for mixed logit models), including the need to make feature-specific assumptions about how heterogeneity is distributed.

By increasing the flexibility in how unobserved heterogeneity can be modeled and eliminating the need for restricting functional form assumptions, MAPL models significantly narrow the gap between the theoretical possibilities and practical applications of incorporating unobserved heterogeneity into discrete choice models. MAPL models also facilitate the incorporation of machine learning methods into discrete choice modeling, building on recent literature connecting these fields \citep{VanCranenburgh2022}. By incorporating unobserved heterogeneity into machine learning modeling, MAPL models can achieve superior predictive performance compared to state-of-the-art neural network models and achieve at least as good performance as correctly-specified mixed logit models.

The structure of this paper is as follows: \Cref{sec:LitReview} reviews traditional discrete choice modeling paradigms and their connections to machine learning. \Cref{sec:MAPLModel} presents a formal description of the MAPL model. \Cref{sec:SimulationExperiment} presents a simulation experiment where the predictive performance of a MAPL model is compared against that of different mixed logit models and neural network models. Lastly, \Cref{sec:Conclusion} summarizes the paper's findings and directions for future work.

\section{Literature Review and Contributions}\label{sec:LitReview}

Discrete choice modeling has a rich history lasting over one hundred years \citep{McFadden2000Nobel} and is leveraged in a wide variety of fields \citep{Train2009,Haghani2021}. Common discrete choice models include the multinomial logit (MNL), nested logit (NL), and mixed logit (MXL) models \citep{Train2009}. The mixed logit model is particularly notable due to its flexibility and theoretical capabilities in terms of modeling unobserved heterogeneity \citep{McFadden2000}. In practice, however, modelers often make several simplifying assumptions when specifying mixed logit models, including assuming independent mixing distributions, convenient mixing distributions (e.g., normal or log-normal), or preference homogeneity for a subset of product features \citep{Forsythe2023,Helveston2015,Philip2021,Revelt1998, Guo2021,Kavalec1999}. Depending on the software used, these assumptions are not necessarily made by choice as not all software capable of estimating mixed logit models support alternative assumptions.

Recent work has sought to improve some of these restrictions by proposing more flexible mixing distributions \citep{Fosgerau2013,Train2016, Krueger2020}, but more flexible mixing distributions can lead to longer estimation times \citep{Bansal2018}. Researchers have developed more efficient estimation packages \citep{Arteaga2022,Helveston2023,Molloy2021}, but the simulation requirements for equally-precise mixed logit estimation can still scale considerably with the number of features with random taste heterogeneity, larger data sizes \cite[Table 5]{Czajkowski2019}, and in contexts where a model must be estimated multiple times (e.g., multi-start algorithms to search for global optima).

In more recent years, researchers have begun integrating discrete choice modeling with machine learning techniques \citep{VanCranenburgh2022}. For instance, Nobel laureate Daniel McFadden notes in his Nobel lecture that the latent class logit can be represented as an artificial neural network \citep[p.342]{McFadden2000Nobel}. Machine learning models in the context of discrete choice models generally offer superior predictive capability \citep{Salas2022} and require less time to estimate \citep{Wang2021,Garcia-Garcia2022}, but are often not immediately interpretable \citep[Sec. 3.1.2]{VanCranenburgh2022}. Several recent works combine traditional discrete choice modeling techniques with machine learning to yield flexible and interpretable models. Studies such as \citet{Wong2021,Arkoudi2023,Han2022,Sifringer2020} integrate the multinomial logit with neural networks in a variety of ways to relax functional form assumptions while still providing interpretable parameters. \citet{Sifringer2020} additionally focuses on combining nested logit models with neural networks. Even latent class logit models have been combined with neural networks to improve model flexibility \citep{Lahoz2023}. Currently, the integration of machine learning techniques with mixed logit modeling techniques is missing from the literature. 

The MAPL model makes two notable contributions to existing literature. First, it provides a novel framework for modeling unobserved heterogeneity that allows for model inputs to directly predict parameters of alternative-specific distributions of unobserved preference heterogeneity. This structure also removes the need to make assumptions about the functional form for the latent decision model, further relaxing the assumptions required by the modeler. Second, the model allows for a novel extension of machine learning models that explicitly models continuous, unobserved consumer heterogeneity akin to the mixed logit. In making this integration, MAPL models have the capacity to continue improving by integrating future innovations that emerge from the field of machine learning. These contributions have the opportunity to greatly impact discrete choice modeling through alternative estimation schemes, expanded flexibility in modeling heterogeneous preference, and improved prediction performance.

\section{The MAPL Model}\label{sec:MAPLModel}

The MAPL model is distinguished by its conception of unobserved heterogeneity. Traditional mixed logit models require the modeler specify feature-specific heterogeneity distributions, which may or may not be accurate assumptions. In contrast, MAPL models directly relate model inputs to distributional parameters of aggregate observables-related preference (see Figure \ref{fig:SimpleHeterogeneityFigure}). This frees the modeler from having to make feature-specific assumptions about heterogeneity and provides greater flexibility in fitting models with heterogeneity in preferences.

\begin{figure}[H]
    \centering
    \begin{tikzpicture}
        \draw[white](0,0)--(0,0);
        \begin{scope}[shift={(0,0)}]
        
            \draw[->] (-1.5,0) -- (1.5,0) node[right] {};

            \def\stddev{0.35}
        
            \draw[smooth, thick, domain=-1:1] plot (\x, {exp(-\x*\x/(2*\stddev*\stddev))/(sqrt(2*pi)*\stddev)});

        \end{scope}

        \node[black] at (1.75,1) {=};
        
        \node[black] at (4.75,1) {\Huge $\cdot$};
        \node[black] at (5.75,1) {\large $x_{ijt}$};
        \node[black] at (6.75,1) {+};
        \node[black] at (9.25,1) {\Huge $\cdot$};
        \node[black] at (10.25,1) {\large $z_{ijt}$};
        \node[black, align=center] at (8,-.45) {$z$-related\\preference};
        \node[black, align=center] at (3.5,-.45) {$x$-related\\preference};
        
        \node[black, align=center] at (0,-.45) {Heterogeneous\\preference};
        \begin{scope}[shift={(8,0)}]
        
            \draw[->] (-1.5,0) -- (1.5,0) node[right] {};

            \def\stddev{0.25}
        
            \draw[smooth, thick, domain=-1:1] plot (\x, {exp(-\x*\x/(2*\stddev*\stddev))/(sqrt(2*pi)*\stddev)});

        \end{scope}

        \begin{scope}[shift={(3.5,0)}]
        
            \draw[->] (-1.5,0) -- (1.5,0) node[right] {};

            \def\stddev{0.2}
        
            \draw[smooth, thick, domain=-1:1] plot (\x, {exp(-\x*\x/(2*\stddev*\stddev))/(sqrt(2*pi)*\stddev)});

        \end{scope}

        \draw[red, <->, thick] (3.75, 2.1) .. controls (4.4,3) and (6.875,3) .. (7.25, 2.1);
        \node[red, align=center] at (5.75,3.5) {\large Mixed logit models estimate parameters of\\ \large constituent feature-specific mixing distributions};
        \draw[blue, ->, thick] (5.75, .65) .. controls (5.5,-2) and (5,-2) .. (1, -1);
        \draw[blue, -, thick] (10.25, .65) .. controls (10,-2) and (9.5,-2) .. (4, -1.62);
        \node[blue, align=center] at (5.75,-3) {\large MAPL models directly relate model inputs to parameters of\\ \large alternative-specific distributions of aggregate preference heterogeneity};
    \end{tikzpicture}
    \caption{Conceptual diagram comparing how heterogenity is modeled in MAPL versus mixed logit models. Whereas mixed logit models estimate parameters of feature-specific mixing distributions, MAPL models predict distribution parameters for aggregate preferences based on model inputs.}
    \label{fig:SimpleHeterogeneityFigure}
\end{figure}

\subsection{Model Structure}\label{ssec:MAPL-Structure}

Specifying a MAPL model centers around three modeling decisions: the choice preference framework, the aggregate preference distribution specification, and the preference distribution parameter estimator. 

\begin{figure}[H]
    \centering
    \begin{tikzpicture}[
        neuron/.style={circle, draw, minimum size=1cm},
        layer/.style={anchor=center},
        arrow/.style={-Stealth}
    ]
    
    \node[neuron] (input1) at (0, 6) {$x_{1}$};
    \node[neuron] (input2) at (0, 4) {$x_{2}$};
    \node[yshift=-.9cm] at (input2) {$\vdots$};
    \node[neuron] (input3) at (0, 2) {$x_{k}$};
    \node[layer, above=of input1, yshift=-1cm, align=center] {\large Product\\ \large Attributes};
    \foreach \j in {1,2,3} {
        \node[neuron, right=.75cm of input2, yshift=(\j-2)*2cm] (hidden1\j) {};
    }
    \node[yshift=-.9cm] at (input1) {$\vdots$};
    \node[layer, above=of hidden13, yshift=-1cm, align=center] {\large Layer\\\large 1};
    
    \foreach \j in {1,2,3} {
        \node[neuron, right=.75cm of hidden12, yshift=(\j-2)*2cm] (hidden2\j) {};
    }
    \node[yshift=-.9cm] at (hidden23) {$\vdots$};
    \node[yshift=-2.9cm] at (hidden23) {$\vdots$};
    \node[layer, above=of hidden23, yshift=-1cm, align=center] {\large Layer \\ \large N};
    
    \foreach \i in {1,2,3} {
        \foreach \j in {1,2,3} {
            \draw[arrow] (input\i) -- (hidden1\j);
        }
    }
    
    \node[xshift=.9cm] at (hidden13) (topdots) {$\cdots$};
    \node[yshift=-2cm] at (topdots) {$\cdots$};
    \node[yshift=-4cm] at (topdots) {$\cdots$};
    \node[layer, above=of topdots, align=center, yshift=1.5cm] {\Large \underline{Machine Learning}\\ \Large \underline{Model}};
    \node[layer, above=of topdots, align=center, yshift=0.5cm] {Train ML model to learn parameters\\of heterogeneity distribution};
    
    \node[neuron, right=.75cm of hidden22, yshift=1cm] (output1) {$\theta_{1}$};
    \node[neuron, right=.75cm of hidden22, yshift=-1cm] (output2) {$\theta_{m}$};
    \node[layer, above=of output1, yshift=-1cm] {\large Output};
    \node[yshift=-.9cm] at (output1) {$\vdots$};
    
    \foreach \i in {1,2,3} {
        \foreach \m in {1,2} {
            \draw[arrow] (hidden2\i) -- (output\m);
        }
    }

    \draw[arrows={-Stealth[inset=0pt, length=.25cm, width=.25cm]}, thick] (6, 4) -- (7, 4);

    \begin{scope}[shift={(8.5,2)}, scale=1.5]
        
            \draw[->] (-1,0) -- (1,0) node[right] {};

            \def\stddev{0.15}
        
            \draw[smooth, thick, domain=-1:1] plot (\x, {exp(-\x*\x/(2*\stddev*\stddev))/(sqrt(2*pi)*\stddev)});

    \end{scope}
    \node[layer, above=of topdots, align=center, yshift=1.5cm, xshift=5.9cm] {\Large \underline{Preference}\\\Large \underline{Distribution}};
    \node[layer, above=of topdots, align=center, yshift=-0.2cm, xshift=5.9cm] {Simulate\\preference distribution\\given predicted\\distributional parameters};
    \node[black, align=center] at (8.5,1.6) {\large Utility ($\nu$)};
    \draw[arrows={-Stealth[inset=0pt, length=.25cm, width=.25cm]}, thick] (10, 4) -- (11, 4);

    \node[layer, above=of topdots, align=center, yshift=1.5cm, xshift=10.75cm] {\Large \underline{Simulated}\\\Large \underline{Log-Likelihood}};
    \node[layer, above=of topdots, align=center, yshift=0.5cm, xshift=10.75cm] {Estimate objective function\\via integration by simulation};
    \node[layer, above=of topdots, align=left, yshift=-3cm, xshift=10.75cm] {\large SLL$=\displaystyle\sum_{n=1}^N \displaystyle\sum_{j=1}^{J} d_{nj}\ln{\hat{P}_{nj}} $};
    \node[layer, above=of topdots, align=left, yshift=-3.5cm, xshift=9.2cm] {\large where};
    \node[layer, above=of topdots, align=left, yshift=-5cm, xshift=10.5cm] {\large $\displaystyle\hat{P}_{nj}=\frac{1}{R}\displaystyle\sum_{r=1}^R L_{nj}\paren{\vectG{\nu}_r}$};
    
\end{tikzpicture}
    \caption{Conceptual diagram of a utility-maximizing MAPL model leveraging a neural network as a parameter estimator. $K$ product attributes ($x_k$) that are passed through a machine learning model (e.g., neural network) that predicts $M$ distributional parameters ($\theta_m$). Those distributional parameters are used to simulate distributions of utility ($\nu$). Separate, independent distributions of utility are produced for each product $j$ among $J$ products. Preference values across all products ($\vectG{\nu}_r$; e.g., utilities) are simulated $R$ times to integrate by simulation the probability linking function ($L$; e.g., logit) over the product preference distributions. Finally, the likelihood function (SLL) is calculated a lá mixed logit estimation \citep[Ch.6]{Train2009}.}
    \label{fig:HeterogeneityFigure}
\end{figure}

\subsubsection{Choice Preference Framework}

The choice preference framework dictates the modeled decision-making paradigm, which is chosen based on the modeler's belief regarding individuals' decision-making processes. MAPL models assume there is some continuously distributed latent quantity associated with each choice alternative that can be transformed to estimate choice probabilities (e.g., via the logit function). As this is a specific functional form assumption not required by MAPL, we prefer to use the generic term ``preference" over the commonly-used ``utility.'' While a utility maximization paramdigm is commonly assumed in many domains, others prefer a regret minimization framework \citep[p.79]{Chorus2012}. Moreover, future modelers might consider alternative conceptions of latent preferences that a MAPL model could incorporate. Still, it is incumbent upon the modeler to leverage institutional knowledge or empirical evidence to decide upon the relationship between preference quantities and choice probability estimates.

\subsubsection{Aggregate Preference Distribution Specification}

The aggregate preference distribution specification is the explicitly assumed distribution for aggregate preference heterogeneity. For instance, assuming a normal distribution is assuming that the aggregate preference for a given product is distributed unimodally and symmetrically across the population of interest. Alternatively, if one chose a mixture of normals, one would be assuming that there is a finite set of subpopulations that have distinct, normally distributed preferences. Aggregate preference distributions are generally assumed to be univariate,\footnote{MAPL models do not preclude assuming multi-dimensional distributions; for instance, one could theoretically model a multi-dimensional distribution of utility and regret if applicable.} but can be any parameterizable distribution with a finite integral, such as a Normal, Lognormal, or Triangular. Moreover, the model can incorporate more flexible non- or semi-parametric distributional forms like those described in \citet{Fosgerau2013,Train2016,Krueger2020}. We demonstrate the use of the semi-parametric distribution from \citet{Fosgerau2013} in our simulation experiment in \Cref{sec:SimulationExperiment}. Modelers should make distributional assumptions based on institutional knowledge or empirical evidence. 

\subsubsection{Preference Distribution Parameter Estimator}

The choice of a parameter estimator relates model inputs to the parameters governing consumer preference. This decision reflects the modeler's belief regarding the complexity associated with transforming model inputs to distributional parameters. This could be as simple as a linear model or as complex as a neural network, which is used in the simulation experiment in \Cref{sec:SimulationExperiment}. Any estimator that can predict scalar values can be incorporated into a MAPL model. Again, this decision should be made based on institutional knowledge or empirical evidence as well as the goals of the modeling exercise (e.g. predictive accuracy).

\subsection{Mathematical Representation of the MAPL Model}\label{ssec:MAPL-Math}

Assuming $n$ total choice alternatives, individual $i$ has chosen an alternative $j$ within choice set $t$ according to some decision-making paradigm. Choice decisions are represented by $y_{ijt}\in \bracket{0,1}$ (\vectA{y} when observations are stacked), and $k$ explanatory variables (e.g., product features, consumer demographics, etc.) are contained within the vector $\vectA{x}_{ijt}\in \mathbb{R}^k$ ($\vectA{X}$ when observations are stacked). These model inputs as well as a vector of $m$ estimable parameters ($\vbeta\in \mathbb{R}^m$) serve as arguments to some function $\vectA{G}\paren{\vbeta, \vectA{X}}\in \mathbb{R}^{s\cdot n}$ that maps $k$ model inputs to $s$ parameters for each observation. The full set of $s\cdot n$ parameters define a vector of preference distributions $\vpsi \sim \vPsi\paren{\vectA{G}\paren{\vbeta, \vectA{X}}}$. Next, some linking function (e.g., the logit function) $\vectA{h}\paren{\vpsi}$ estimates choice probabilities ($\hat{\vectA{p}}$) based on a given vector of preference values. Some measure of the estimated choice probabilities that accounts for heterogeneity ($\hat{\vectA{p}}^*$) is used to compare model predictions to outcomes. A common choice for $\hat{\vectA{p}}^*$ will likely be the expected value of choice probabilities, which can be found by integrating said linking function across preference distributions. Similar to the integration over parameters in the mixed logit framework \citep[Ch.6]{Train2009}, \cref{eq:probability_estimator} shows the relevant integral required to identify the expected choice probabilities for a set of alternatives where $f$ is the appropriate probability density function for the $\vPsi\paren{\vectA{G}\paren{\vbeta, \vectA{X}}}$ vector of distributions. Parameters $\vbeta$ are estimated by solving \cref{eq:minimization_problem} where $\vbeta$ minimizes some measure of fit $d$ (e.g., log-likelihood, KL-divergence).

\begin{equation}\label{eq:probability_estimator}
    \bar{\hat{\vectA{p}}}\paren{\vbeta, \vectA{X}} = \int \vectA{h}\paren{\vpsi}f(\vpsi, \vbeta, \vectA{X})d\vpsi
\end{equation}

\begin{equation}\label{eq:minimization_problem}
    \begin{aligned}
        \min_{\vectG{\beta}} \quad & d(\vectA{y},\hat{\vectA{p}}^*\paren{\vbeta, \vectA{X}}) \\
    \end{aligned}
\end{equation}

\section{Simulation Experiment}\label{sec:SimulationExperiment}

\subsection{Experiment Design}\label{ssec:Design}

To examine the performance and flexibility of MAPL, we conduct a simulation experiment where we simulate sets of choice data according to multiple different data generating processes (DGPs) that follow different variations of mixed logit utility models with different utility structures and different heterogeneity distributions. Our experiment is motivated by the fact that modelers do not know \emph{a priori} the underlying utility functional form nor the distributional form of any unobserved heterogeneity in the true DGP. When using mixed logit, inaccurate assumptions about the functional form or heterogeneity distributions will lead to poorer performance. In contrast, the flexibility of the MAPL model should produce good performance for each DGP without making any assumptions about the utility functional form or feature-level assumptions about preference heterogeneity.

To test this, we employ four different mixed logit DGPs to simulate different sets of choice data. Each DGP has a fixed parameter, $\beta_0$, and two random parameters, $\beta_1$ and $\beta_2$. In DGP 1, $\beta_1$ and $\beta_2$ follow independent normal distributions. DGP 2 is the same as 1 except that the two distributions are correlated. For DGPs 3 and 4, we use independent normal distributions for $\beta_1$ and $\beta_2$, except that DGP 3 includes an additonal interaction parameter between $x_0$ and $x_1$, and DGP 4 includes a nonlinear parameter for $x_1^2$. \Cref{table:dgp} below summarizes each of the DGPs. All models follow a random utility theory decision-making paradigm where the utility is formed by a summation of a portion relating to $x$ features, $\nu_{ijt}$, and an error term, $\epsilon_{ijt}$, representing a Gumbel-distributed error term such that $u_{ijt} = \nu_{ijt} + \epsilon_{ijt}$. 

\begin{table}[h]
\centering
\caption{Summary of data generating processes (DGPs) used in simulation experiment.}
\begin{tabular}{p{3.2cm}m{5.8cm}m{5.6cm}} 
\toprule
\textbf{Scenario} & \textbf{$\nu_{ijt}$} & Heterogeneity specification \\
\midrule
Independent\newline normals & $\beta_{0} x_{0jt} + \beta_{i1} x_{1jt} + \beta_{i2} x_{2jt}$ & $\begin{bmatrix} \beta_{1} \\ \beta_{2} \end{bmatrix} \sim \mathrm{Normal} \left( \begin{bmatrix} \mu_{1} \\ \mu_{2} \end{bmatrix}, \begin{bmatrix} \sigma_1^2 & 0 \\ 0 & \sigma_2^2 \end{bmatrix} \right)$ \\
\midrule
Correlated\newline normals & $\beta_{0} x_{0jt} + \beta_{i1} x_{1jt} + \beta_{i2} x_{2jt}$ & $\begin{bmatrix} \beta_{1} \\ \beta_{2} \end{bmatrix} \sim \mathrm{Normal} \left( \begin{bmatrix} \mu_{1} \\ \mu_{2} \end{bmatrix}, \begin{bmatrix} \sigma_1^2 & \sigma_{12}^2 \\ \sigma_{12}^2 & \sigma_2^2 \end{bmatrix} \right)$ \\
\midrule
Independent normals\newline w/interaction & $\beta_{0} x_{0jt} + \beta_{i1} x_{1jt} + \beta_{i2} x_{2jt} + \beta_{3} x_{0jt} x_{1jt}$ & $\begin{bmatrix} \beta_{1} \\ \beta_{2} \end{bmatrix} \sim \mathrm{Normal} \left( \begin{bmatrix} \mu_{1} \\ \mu_{2} \end{bmatrix}, \begin{bmatrix} \sigma_1^2 & 0 \\ 0 & \sigma_2^2 \end{bmatrix} \right)$ \\
\midrule
Independent normals\newline w/nonlinear term & $\beta_{0} x_{0jt} + \beta_{i1} x_{1jt} + \beta_{i2} x_{2jt} + \beta_{3} x_{1jt}^2$ & $\begin{bmatrix} \beta_{1} \\ \beta_{2} \end{bmatrix} \sim \mathrm{Normal} \left( \begin{bmatrix} \mu_{1} \\ \mu_{2} \end{bmatrix}, \begin{bmatrix} \sigma_1^2 & 0 \\ 0 & \sigma_2^2 \end{bmatrix} \right)$ \\
\bottomrule
\end{tabular}
\label{table:dgp}
\end{table}

Product features (the various $x$ values) in each respective DGP are uniformally distributed on a domain of $[-1, 1]$. With each DGP, we simulate 10,000 individuals making 10 successive choices each from among three choice alternatives. Choice probabilities for each alternative are derived according to traditional random utility theory \citep{Train2009} using simulation to obtain the mean $P_{jt}$ across 1,000 draws of the logit function, $P_{jt} = \exp({\nu_{jt}}) / \sum_{k=1}^{J} \exp({\nu_{kt}})$. The coefficients used in each DGP are as follows: $\beta_{0} = -1$, $\mu_{1} = 1$, $\mu_{2} = 2$, $\sigma_{1} = 1$, $\sigma_{2} = 1.5$, $\sigma_{12} = 0.7$, $\beta_{3} = 2$

For each DGP, we estimate a variety of choice-models specifications that are common among pracitioners: a simple multinomial logit (MNL), a Mixed Logit (MXL) with independent mixing distributions, a Simple Neural Network (NN), a Deep NN, and two forms of MAPL models: one using normal distributions for the aggregate preference heterogeneity distributions and one using the semi-parametric distribution posed by \citet{Fosgerau2013} with 12 distributional parameters. The MAPL models were trained using neural networks with two hidden layers composed of 64 neurons each as well as a normalization and dropout layer after each hidden layer. Importantly, the neural network transforms model inputs for each choice alternative independently and have identical weights. We train each model repeatedly across twenty instantiations of the DGPs using negative log-likelihood as the loss function for training with an 80/20 train/test split and 2,000 epochs per training session. Full model specifications can be found in this paper's replication code available at \url{https://github.com/crforsythe/MAPL-Paper}.

\subsection{Experiment Results}\label{ssec:Results}

\begin{figure}[H]
    \centering
    \includegraphics[width=1\linewidth]{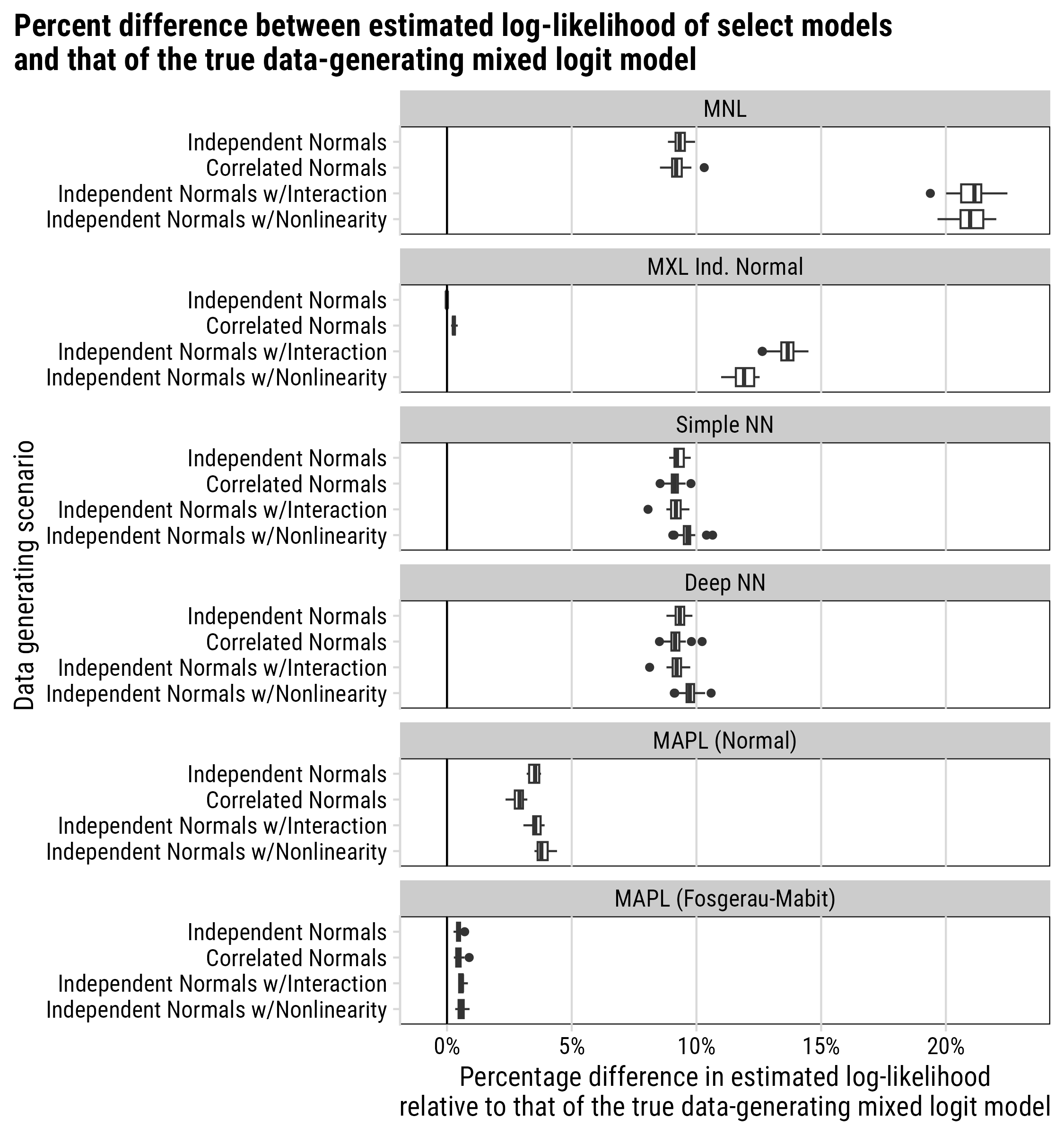}
    \caption{Boxplots of differences in estimated and true log-likelihood values across a variety of model specifications and data generating process. The MAPL model achieves superior performance without needing to assume a utility model form or feature-level heterogeneity distributional forms. The Fosgerau-Mabit aggregate preference heterogeneity distribution out-performs the normal distribution given its greater flexibility.}
    \label{fig:Misspecification-Boxplot}
\end{figure}

\Cref{fig:Misspecification-Boxplot} presents the differences between the estimated log-likelihood values of various model specifications and the true log-likelihood value across all DGPs, where the true log-likelihood is taken as the estimated log-likelihood of a mixed logit model that exactly matches the true underlying DGP. Results are presented as box plots since we conducted the experiment 20 times for each combination of DGP and model. Several observations are of note.

First, we see that models that misspecify unobserved heterogeneity have relatively poor performance. The MNL models (which cannot account for unobserved heterogeneity) have errors of approximately 9\% even with correct utility functional forms and errors above 20\% with incorrect utility functional forms. Likewise, heterogeneity misspecification is all but guaranteed for the traditional machine learning models (Simple NN and Deep NN) as these models are unable to account for unobserved heterogeneity. However, it is notable that these models perform equally well across different utility functional forms, such as the DGPs with interactions and nonlinearities.

Second, the mixed logit model with independent normals performs perfectly (0\% error) in the case where the true DGP is exactly the same, as would be expected. This model also performs very well even when heterogeneity is correlated, suggesting that the common assumption of uncorrelated heterogeneity may be reasonable. However, misspecification in utility form leads to large discrepancies in estimate and true log-likelihood values: 12\% when omitting the non-linear term and 14\% when omitting the interaction term. 

Finally, the results confirm that the MAPL models were robust to all DGP specifications, even though they require no \emph{a priori} specification of utilty. A MAPL model with a simple assumption of aggregate preference heterogeneity being normally-distributed never exeeded more than 4\% error, and this gap shrunk to $<$1\% for all iterations when using the more flexible Fosgerau-Mabit distribution, suggesting that flexible heterogeneity distributions are able to capture the underlying true heterogeneity. This illustrates the significance of incorporating unobserved heterogeneity into machine learning-based modeling approaches.

While the results in \Cref{fig:Misspecification-Boxplot} are promising, it is important to note that this experiment was conducted using a relatively large sample size (10,000 simulated sets of 10 choices, for a total of 100,000 choice observations). While such sizes may be available in some domains, in others sample sizes may be constrained, such as in stated-choice experiments where sizes are often limited by research budgets. To examine the potential impact of sample size on the performance MAPL models, we repeat one of the simulation scenarios using different sample sizes ranging from 500 to 10,000 sets of 10 observations (5,000 to 100,000 choice observations). We use DGP1 where $\beta_1$ and $\beta_2$ follow independent normal distributions for the data simulation and fit a MAPL model with the Fosgerau-Mabit aggregate preference heterogeneity distribution. 

\begin{figure}[H]
    \centering
    \includegraphics[width=0.75\linewidth]{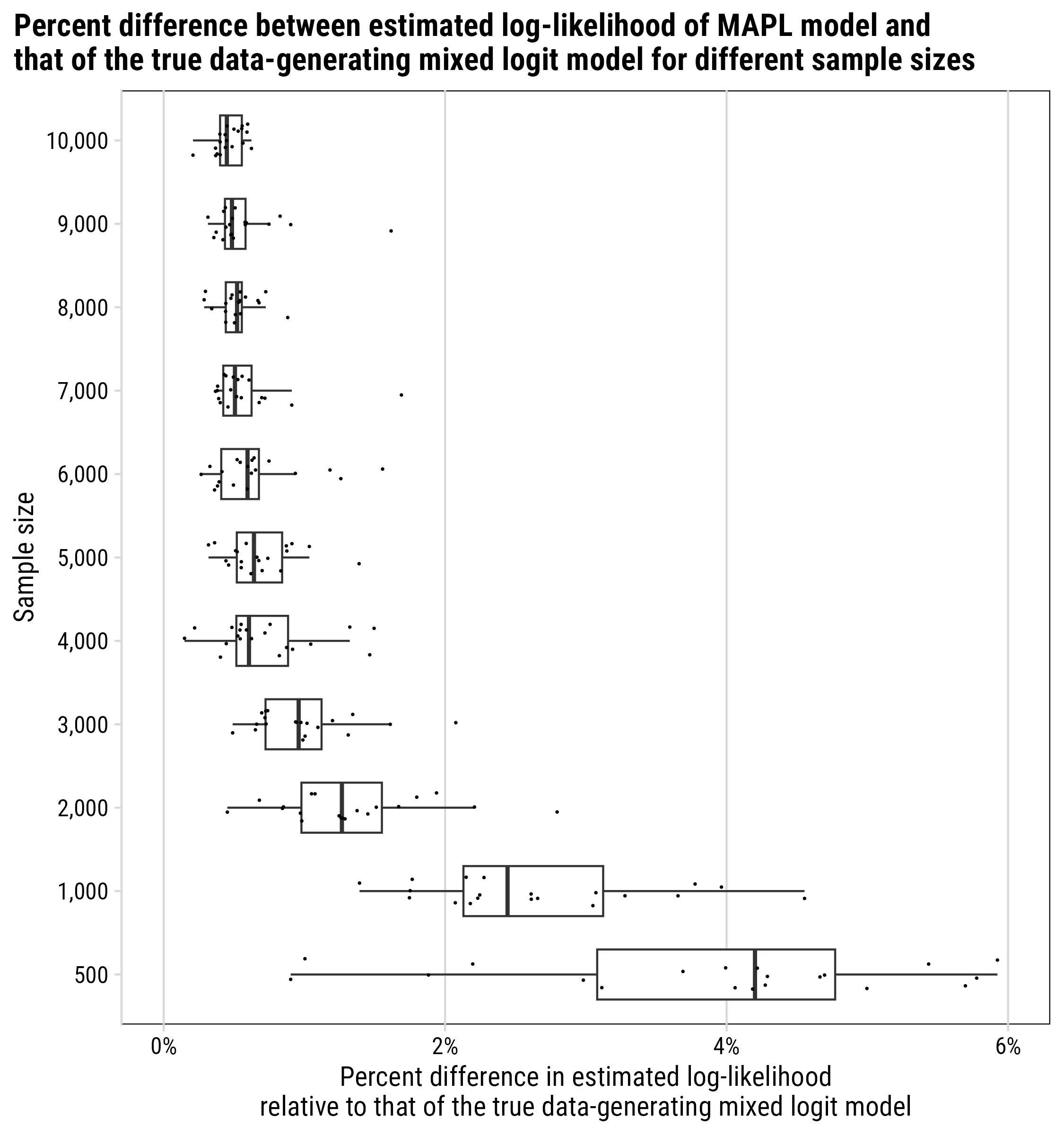}
    \caption{Boxplots of differences in estimated and true log-likelihood values using a MAPL model with the Fosgerau-Mabit aggregate preference heterogeneity distribution and increasing sample sizes. For each sample size, we run 20 independent simulations generating the data using DGP1 as described in \Cref{ssec:Design}. While error increases with lower sample sizes, even at only 500 simulated people, errors never exceed 6\%, which is still significantly below the errors of MNL and NN models with 10,000 simulated people.}
    \label{fig:sample-size}
\end{figure}

\Cref{fig:sample-size} shows the results of this sensitivity analysis, revealing that the performance of MAPL models is constrained by sample size, which is expected for models using neural networks \citep{van2021artificial}. Errors are higher with sample sizes under 1,000 simulated people and then quickly fall to below 2\% and stabilize at under 1\% at 4,000 people and more. However, even at only 500 simulated people, errors never cross above 6\%, which is still significantly below the errors of MNL and NN models with 10,000 simulated people. We also observe that the variation in the 20 independent iterations of the simulation shrinks with increased sample size.

\section{Discussion and Conclusion}\label{sec:Conclusion}

This paper introduces the \textbf{M}ixed \textbf{A}ggregate \textbf{P}reference \textbf{L}ogit (MAPL) model, a novel class of discrete choice models that significantly advances how researchers can incorporate unobserved heterogeneity into choice analysis. By directly characterizing alternative-specific distributions of preference heterogeneity as functions of model inputs, MAPL models offer several advantages over traditional approaches. First, they eliminate the need to specify feature-specific heterogeneity distributions, freeing modelers from potentially restrictive assumptions. Second, they remove the requirement to define a specific functional form for the underlying decision-making process. Instead, MAPL models integrate machine learning techniques to flexibly capture complex relationships between inputs and preference distributions.

Our simulation experiment demonstrates that the flexibility of MAPL models yields superior predictive performance compared to traditional neural networks that cannot account for unobserved heterogeneity. Importantly, a single MAPL model specification performs well across multiple data-generating processes with different utility structures and heterogeneity distributions, without requiring the modeler to correctly identify the true underlying model. This robustness is particularly valuable in real-world applications where the true decision-making process and heterogeneity structure are unknown.

The benefits of MAPL models extend beyond prediction performance. For policy analysts, market researchers, and product designers, MAPL models offer more reliable predictions of choice behavior under varying scenarios, potentially leading to better-informed decisions. Furthermore, MAPL models can serve as diagnostic tools for traditional choice modelers, helping identify potential misspecification in utility functions or heterogeneity distributions. By comparing the performance of a MAPL model using flexible distributions (such as the semi-parametric distribution proposed by \citet{Fosgerau2013}) with that of traditional mixed logit specifications, researchers can detect areas where traditional models may be missing important preference structures.

Despite these advantages, MAPL models do face certain limitations. As shown in our sample size sensitivity analysis, their performance depends on having sufficient data. However, MAPL models with smaller sample sizes still out-perform other models even when those models are estimated with substantially larger sample sizes. Additionally, the current formulation focuses on alternative-specific distributions, which may not capture all forms of preference heterogeneity, particularly those that manifest as correlations across alternatives.

Several promising directions for future research emerge from this work. First, developing robust techniques for interpreting MAPL models represents a crucial area for further work. While MAPL models offer excellent predictive capabilities, extracting meaningful behavioral insights remains challenging. We aim to adapt interpretability methods from machine learning literature \citep{Linardatos2021,van2021artificial} to help researchers understand the patterns of heterogeneity captured by these models. Of particular interest will be recovering economically meaningful quantities such as willingness-to-pay distributions, elasticities, and marginal effects from trained MAPL models. Second, future research should explore theoretical connections between MAPL models and traditional choice models, particularly establishing conditions under which MAPL models can exactly recover traditional specifications. This work would help position MAPL models within the broader discrete choice literature and clarify when they offer meaningful advantages over existing approaches. Third, computational efficiency remains an important consideration. While MAPL models reduce the dimensionality of integration compared to traditional mixed logit, the training process for complex neural network components can be computationally intensive. Research into more efficient training algorithms and model architectures could further enhance the practical utility of MAPL models. Finally, extensions of the MAPL framework to incorporate dynamic choice behavior, reference dependence, or other behavioral economic phenomena represent promising directions for expanding the approach. The flexibility of the MAPL modeling framework makes it well-suited to capturing these more complex choice processes.

In conclusion, MAPL models offer a powerful new approach to modeling unobserved heterogeneity in discrete choice analysis by combining the strengths of traditional econometric models with the flexibility of machine learning methods. They address key limitations in existing approaches and open new possibilities for understanding and predicting choice behavior across a wide range of applications.

\section{Glossary}

\noindent\textbf{DGP} Data Generating Process
\newline

\noindent\textbf{MAPL} Mixed Aggregate Preference Logit
\newline

\noindent\textbf{MNL} Multinomial Logit
\newline

\noindent\textbf{MXL} Mixed Logit
\newline

\noindent\textbf{NL} Nested Logit
\newline

\noindent\textbf{NN} Neural Network

\vspace{15pt}

\section{CRediT authorship contribution statement}

\textbf{Connor R. Forsythe:} Conceptualization, Formal analysis, Investigation, Methodology, Software, Supervision, Validation, Writing – original draft, Writing – review \& editing. \textbf{Cristian Arteaga:} Data curation, Formal analysis, Investigation, Software, Supervision, Validation, Writing – review \& editing. \textbf{John Paul Helveston:} Data curation, Formal analysis, Investigation, Software, Visualization, Writing – review \& editing.

\vspace{15pt}

\section{Acknowledgments}

We would like to acknowledge Dr. Jeremy Michalek and Dr. Kate Whitefoot for their feedback that helped guide the direction of inquiry for this project.

\setlength{\bibsep}{0pt}
\bibliographystyle{aer}
\bibliography{bib}

@article{McFadden2000,
author = {McFadden, Daniel and Train, Kenneth},
doi = {10.1002/1099-1255(200009/10)15:5<447::aid-jae570>3.0.co;2-1},
issn = {08837252},
journal = {Journal of Applied Econometrics},
number = {5},
pages = {447--470},
title = {{Mixed MNL models for discrete response}},
volume = {15},
year = {2000}
}

@article{Forsythe2023,
author = {Forsythe, Connor R. and Gillingham, Kenneth T. and Michalek, Jeremy J. and Whitefoot, Kate S.},
doi = {10.1073/pnas.2219396120},
issn = {10916490},
journal = {Proceedings of the National Academy of Sciences of the United States of America},
number = {23},
pages = {1--7},
pmid = {37252977},
title = {{Technology advancement is driving electric vehicle adoption}},
volume = {120},
year = {2023}
}

@article{Helveston2015,
author = {Helveston, John Paul and Liu, Yimin and Feit, Elea Mc Donnell and Fuchs, Erica and Klampfl, Erica and Michalek, Jeremy J.},
doi = {10.1016/j.tra.2015.01.002},
issn = {09658564},
journal = {Transportation Research Part A: Policy and Practice},
pages = {96--112},
publisher = {Elsevier Ltd},
title = {{Will subsidies drive electric vehicle adoption? Measuring consumer preferences in the U.S. and China}},
url = {http://dx.doi.org/10.1016/j.tra.2015.01.002},
volume = {73},
year = {2015}
}

@book{Train2009,
author = {Train, Kenneth},
edition = {2},
publisher = {Cambridge University Press},
title = {{Discrete Choice Methods with Simulation}},
year = {2009}
}

@article{Fosgerau2013,
author = {Fosgerau, Mogens and Mabit, Stefan L.},
doi = {10.1016/j.econlet.2013.03.050},
issn = {01651765},
journal = {Economics Letters},
number = {2},
pages = {206--210},
publisher = {Elsevier B.V.},
title = {{Easy and flexible mixture distributions}},
url = {http://dx.doi.org/10.1016/j.econlet.2013.03.050},
volume = {120},
year = {2013}
}

@article{Train2016,
author = {Train, Kenneth},
doi = {10.1016/j.jocm.2016.07.004},
issn = {17555345},
journal = {Journal of Choice Modelling},
pages = {40--53},
publisher = {Elsevier},
title = {{Mixed logit with a flexible mixing distribution}},
url = {http://dx.doi.org/10.1016/j.jocm.2016.07.004},
volume = {19},
year = {2016}
}

@article{Krueger2020,
author = {Krueger, Rico and Rashidi, Taha H. and Vij, Akshay},
doi = {10.1016/j.jocm.2020.100229},
issn = {17555345},
journal = {Journal of Choice Modelling},
number = {May},
pages = {100229},
publisher = {Elsevier Ltd},
title = {{A Dirichlet process mixture model of discrete choice: Comparisons and a case study on preferences for shared automated vehicles}},
url = {https://doi.org/10.1016/j.jocm.2020.100229},
volume = {36},
year = {2020}
}

@article{Haghani2021,
author = {Haghani, Milad and Bliemer, Michiel C.J. and Hensher, David A.},
doi = {10.1016/j.jocm.2021.100303},
issn = {17555345},
journal = {Journal of Choice Modelling},
number = {June},
pages = {100303},
publisher = {Elsevier Ltd},
title = {{The landscape of econometric discrete choice modelling research}},
url = {https://doi.org/10.1016/j.jocm.2021.100303},
volume = {40},
year = {2021}
}

@article{Chorus2012,
author = {Chorus, Caspar},
doi = {10.1080/01441647.2011.609947},
issn = {01441647},
journal = {Transport Reviews},
number = {1},
pages = {75--92},
title = {{Random Regret Minimization: An Overview of Model Properties and Empirical Evidence}},
volume = {32},
year = {2012}
}

@article{Wang2021,
archivePrefix = {arXiv},
arxivId = {2102.01130},
author = {Wang, Shenhao and Mo, Baichuan and Hess, Stephane and Zhao, Jinhua},
eprint = {2102.01130},
title = {{Comparing hundreds of machine learning classifiers and discrete choice models in predicting travel behavior: an empirical benchmark}},
url = {http://arxiv.org/abs/2102.01130},
year = {2021}
}

@article{Revelt1998,
author = {Revelt, David and Train, Kenneth},
doi = {10.1162/003465398557735},
issn = {00346535},
journal = {Review of Economics and Statistics},
number = {4},
pages = {647--657},
title = {{Mixed logit with repeated choices: Households' choices of appliance efficiency level}},
volume = {80},
year = {1998}
}

@misc{McFadden2000Nobel,
  author = {McFadden, Daniel L.},
  title = {Prize Lecture: Economic Choices},
  year = {2000},
  howpublished = {\url{https://www.nobelprize.org/prizes/economic-sciences/2000/mcfadden/lecture/}},
  note = {Accessed: 2024-01-21},
  publisher = {Nobel Prize Outreach AB}
}

@article{Philip2021,
author = {Philip, Thara and Whitehead, Jake and Prato, Carlo G.},
doi = {10.1016/j.tra.2022.11.015},
issn = {09658564},
journal = {Transportation Research Part A: Policy and Practice},
month = {jan},
number = {December 2022},
pages = {103555},
publisher = {Elsevier Ltd},
title = {{Adoption of electric vehicles in a laggard, car-dependent nation: Investigating the potential influence of V2G and broader energy benefits on adoption}},
url = {https://doi.org/10.1016/j.tra.2022.11.015 https://linkinghub.elsevier.com/retrieve/pii/S0965856422002981},
volume = {167},
year = {2023}
}

@article{Guo2021,
author = {Guo, Yujie and Zhang, Yu},
doi = {10.1016/j.trd.2021.102991},
issn = {13619209},
journal = {Transportation Research Part D: Transport and Environment},
number = {August},
pages = {102991},
publisher = {Elsevier Ltd},
title = {{Understanding factors influencing shared e-scooter usage and its impact on auto mode substitution}},
url = {https://doi.org/10.1016/j.trd.2021.102991},
volume = {99},
year = {2021}
}

@article{Kavalec1999,
author = {Kavalec, Chris},
doi = {10.5547/ISSN0195-6574-EJ-Vol20-No3-5},
issn = {01956574},
journal = {Energy Journal},
number = {3},
pages = {123--138},
title = {{Vehicle choice in an aging population: Some insights from a stated preference survey for California}},
volume = {20},
year = {1999}
}

@article{Arteaga2022,
author = {Arteaga, Cristian and Park, Jee Woong and Beeramoole, Prithvi Bhat and Paz, Alexander},
doi = {10.1016/j.jocm.2021.100339},
issn = {17555345},
journal = {Journal of Choice Modelling},
number = {September 2021},
pages = {100339},
publisher = {Elsevier Ltd},
title = {{xlogit: An open-source Python package for GPU-accelerated estimation of Mixed Logit models}},
url = {https://doi.org/10.1016/j.jocm.2021.100339},
volume = {42},
year = {2022}
}

@article{Czajkowski2019,
author = {Czajkowski, Miko{\l}aj and Budzi{\'{n}}ski, Wiktor},
doi = {10.1016/j.jocm.2019.04.003},
issn = {17555345},
journal = {Journal of Choice Modelling},
number = {April},
pages = {73--85},
title = {{Simulation error in maximum likelihood estimation of discrete choice models}},
volume = {31},
year = {2019}
}

@article{VanCranenburgh2022,
archivePrefix = {arXiv},
arxivId = {2101.11948},
author = {van Cranenburgh, Sander and Wang, Shenhao and Vij, Akshay and Pereira, Francisco and Walker, Joan},
doi = {10.1016/j.jocm.2021.100340},
eprint = {2101.11948},
issn = {17555345},
journal = {Journal of Choice Modelling},
number = {December 2021},
pages = {100340},
publisher = {Elsevier Ltd},
title = {{Choice modelling in the age of machine learning - Discussion paper}},
url = {https://doi.org/10.1016/j.jocm.2021.100340},
volume = {42},
year = {2022}
}

@article{Bansal2018,
author = {Bansal, Prateek and Daziano, Ricardo A. and Achtnicht, Martin},
doi = {10.1016/j.jocm.2017.10.002},
issn = {17555345},
journal = {Journal of Choice Modelling},
number = {September 2017},
pages = {97--113},
publisher = {Elsevier Ltd},
title = {{Comparison of parametric and semiparametric representations of unobserved preference heterogeneity in logit models}},
url = {https://doi.org/10.1016/j.jocm.2017.10.002},
volume = {27},
year = {2018}
}

@article{Linardatos2021,
author = {Linardatos, Pantelis and Papastefanopoulos, Vasilis and Kotsiantis, Sotiris},
doi = {10.3390/e23010018},
issn = {10994300},
journal = {Entropy},
number = {1},
pages = {1--45},
title = {{Explainable ai: A review of machine learning interpretability methods}},
volume = {23},
year = {2021}
}

@article{Molloy2021,
author = {Molloy, Joseph and Becker, Felix and Schmid, Basil and Axhausen, Kay W.},
doi = {10.1016/j.jocm.2021.100284},
issn = {17555345},
journal = {Journal of Choice Modelling},
number = {April 2020},
pages = {100284},
publisher = {Elsevier Ltd},
title = {{mixl: An open-source R package for estimating complex choice models on large datasets}},
url = {https://doi.org/10.1016/j.jocm.2021.100284},
volume = {39},
year = {2021}
}

@article{Helveston2023,
    author = {John Paul Helveston},
    title = {logitr: Fast Estimation of Multinomial and Mixed Logit Models with Preference Space and Willingness to Pay Space Utility Parameterizations},
    journal={Journal of Statistical Software},
    url = {https://jhelvy.github.io/logitr/},
    year = {2023},
    volume = {105},
    number = {10},
    pages = {1--37},
    doi = {10.18637/jss.v105.i10}
}

@article{Wong2021,
archivePrefix = {arXiv},
arxivId = {1912.10058},
author = {Wong, Melvin and Farooq, Bilal},
doi = {10.1016/j.trc.2021.103050},
eprint = {1912.10058},
issn = {0968090X},
journal = {Transportation Research Part C: Emerging Technologies},
number = {January},
pages = {103050},
publisher = {Elsevier Ltd},
title = {{ResLogit: A residual neural network logit model for data-driven choice modelling}},
url = {https://doi.org/10.1016/j.trc.2021.103050},
volume = {126},
year = {2021}
}

@article{Arkoudi2023,
archivePrefix = {arXiv},
arxivId = {2109.12042},
author = {Arkoudi, Ioanna and Krueger, Rico and Azevedo, Carlos Lima and Pereira, Francisco C.},
doi = {10.1016/j.trb.2023.102783},
eprint = {2109.12042},
issn = {01912615},
journal = {Transportation Research Part B: Methodological},
number = {October 2021},
pages = {102783},
publisher = {Elsevier Ltd},
title = {{Combining discrete choice models and neural networks through embeddings: Formulation, interpretability and performance}},
url = {https://doi.org/10.1016/j.trb.2023.102783},
volume = {175},
year = {2023}
}

@article{Garcia-Garcia2022,
author = {Garc{\'{i}}a-Garc{\'{i}}a, Jos{\'{e}} Carlos and Garc{\'{i}}a-R{\'{o}}denas, Ricardo and L{\'{o}}pez-G{\'{o}}mez, Julio Alberto and Mart{\'{i}}n-Baos, Jos{\'{e}} {\'{A}}ngel},
doi = {10.1016/j.trpro.2022.02.047},
issn = {23521465},
journal = {Transportation Research Procedia},
number = {Ewgt 2021},
pages = {374--382},
title = {{A comparative study of machine learning, deep neural networks and random utility maximization models for travel mode choice modelling}},
volume = {62},
year = {2022}
}

@article{Salas2022,
author = {Salas, Patricio and {De la Fuente}, Rodrigo and Astroza, Sebastian and Carrasco, Juan Antonio},
doi = {10.1016/j.eswa.2021.116253},
issn = {09574174},
journal = {Expert Systems with Applications},
number = {November 2021},
pages = {116253},
publisher = {Elsevier Ltd},
title = {{A systematic comparative evaluation of machine learning classifiers and discrete choice models for travel mode choice in the presence of response heterogeneity}},
url = {https://doi.org/10.1016/j.eswa.2021.116253},
volume = {193},
year = {2022}
}

@article{Han2022,
author = {Han, Yafei and Pereira, Francisco Camara and Ben-Akiva, Moshe and Zegras, Christopher},
doi = {10.1016/j.trb.2022.07.001},
issn = {01912615},
journal = {Transportation Research Part B: Methodological},
number = {July},
pages = {166--186},
publisher = {Elsevier Ltd},
title = {{A neural-embedded discrete choice model: Learning taste representation with strengthened interpretability}},
url = {https://doi.org/10.1016/j.trb.2022.07.001},
volume = {163},
year = {2022}
}

@article{Sifringer2020,
archivePrefix = {arXiv},
arxivId = {1812.09747},
author = {Sifringer, Brian and Lurkin, Virginie and Alahi, Alexandre},
doi = {10.1016/j.trb.2020.08.006},
eprint = {1812.09747},
issn = {01912615},
journal = {Transportation Research Part B: Methodological},
pages = {236--261},
title = {{Enhancing discrete choice models with representation learning}},
volume = {140},
year = {2020}
}

@article{Lahoz2023,
archivePrefix = {arXiv},
arxivId = {2302.09871},
author = {Lahoz, Lorena Torres and Pereira, Francisco Camara and Sfeir, Georges and Arkoudi, Ioanna and Monteiro, Mayara Moraes and Azevedo, Carlos Lima},
doi = {10.1016/j.jocm.2023.100452},
eprint = {2302.09871},
issn = {17555345},
journal = {Journal of Choice Modelling},
number = {February},
pages = {100452},
publisher = {Elsevier Ltd},
title = {{Attitudes and Latent Class Choice Models using Machine Learning}},
url = {https://doi.org/10.1016/j.jocm.2023.100452},
volume = {49},
year = {2023}
}

@article{van2021artificial,
  title={An artificial neural network based method to uncover the value-of-travel-time distribution},
  author={van Cranenburgh, Sander and Kouwenhoven, Marco},
  journal={Transportation},
  volume={48},
  number={5},
  pages={2545--2583},
  year={2021},
  publisher={Springer}
}

\newpage
\appendix
\section*{Supplementary Information}\label{sec:SI}

\section{A Special Case of Mixed Logit as a MAPL Model}\label{ssec:MAPL-MixedLogit}

Traditional models of continuously-distributed unobserved consumer preferences (e.g., the mixed logit model) assume that consumer preferences are jointly and continuously distributed across $k$ dimensions, where $k$ is the number of features (i.e., model inputs) with random preference heterogeneity. Modelers then must identify estimated log-likelihood values by integrating the choice probability function over identified distributions. The following example will show how a parsimonious mixed logit model can be framed as a MAPL model.

Consider a parsimonious utility-based mixed logit model \citep[Ch. 6]{Train2009}, where individual $i$ chooses an alternative $j$ within choice set $t$ that maximizes their utility ($u$). Utility is assumed to be composed of observables-associated utility ($\nu_{ijt}$) and a Gumbel-distributed error term ($\epsilon_{ijt}$). We assume utility is linear in its inputs ($w_{ijt}$, $x_{ijt}$, and $z_{ijt}$; see \cref{eq:example_utility_eq}), a single fixed coefficient ($\beta_0$), and two heterogeneous coefficients ($\beta_1$ and $\beta_2$) that are independent and normally distributed (\cref{eq:example_dist}). Mixed logit models estimate parameters by simulating a set of individual-level parameters for each random coefficient, calculating a simulated log-likelihood value \citep{Revelt1998}, and using a nonlinear program solver to estimate constituent model parameters (e.g., $\beta_0$, $\mu_1$, $\mu_2$, $\sigma^2_1$, $\sigma^2_2$). 

\begin{equation}\label{eq:example_utility_eq}
    u_{ijt} = \nu_{ijt}+\epsilon_{ijt} = \beta_0 w_{ijt} + \beta_{i1}x_{ijt}+\beta_{i2}z_{ijt}+\epsilon_{ijt}
\end{equation}

\begin{equation}\label{eq:example_dist}
    \begin{split}
        \bmat
            \beta_{1} \\
            \beta_{2} \\
        \emat
        &\sim
        \up{Normal}\paren{\bmat \mu_1 \\ \mu_2 \emat, \bmat \sigma_1^2, 0 \\ 0, \sigma_2^2 \emat}
    \end{split}
\end{equation}

One issue with this process is that the number of dimensions one must integrate over increases with the number of random parameters. The MAPL model, while making identical assumptions, characterizes preference heterogeneity by assuming a distribution of \emph{aggregate} preference and relating model inputs to distributional parameters of aggregate alternative-specific preference heterogeneity. This reduces the number of dimensions over which one must integrate by $\frac{1}{k}$ (e.g., $\frac{1}{2}$ in this example). Figure \ref{fig:SimpleHeterogeneityFigure} illustrates this distinction between how mixed logit and MAPL models handle unobserved heterogeneity.

Inherent to a MAPL model is the specification of a distribution for aggregate observables-associated preferences (e.g., utility, $u$). For utility-maximizing MAPL models, aggregate observable-associated heterogeneous utility values are assumed to be continuously distributed, and the parameters of this distribution are determined by a set of model parameters and inputs. Given this example's data-generating process, we can choose an appropriate distribution of observables-associated utility. Since any linear combination of normally distributed random variables is itself normally distributed, we can comfortably assume that the observables-associated portion of utility ($\beta_0 w_{ijt}+\beta_{i1}x_{ijt}+\beta_{i2}z_{ijt}$) is itself normally distributed. Further, we can determine how $\beta_0$ and the distributional parameters associated with $\beta_1$ and $\beta_2$ influence the distribution of $\nu$ (see \cref{eq:example_dist_convolution}). 

\begin{equation}\label{eq:example_dist_convolution}
    \begin{split}
        \nu_{ijt} = \beta_0 w_{ijt} + \beta_1 x_{ijt}+\beta_2 z_{ijt}
        &\sim
        \up{Normal}\paren{\mu_{ijt}, \sigma^2_{ijt}} \\
        \mu_{ijt} &= \beta_0 w_{ijt} + \mu_1 x_{ijt}+\mu_2 z_{ijt} \\
        \sigma^2_{ijt} &= \sigma^2_1 x^2_{ijt}+\sigma^2_2 z^2_{ijt}
    \end{split}
\end{equation}

This shift in perspective from constituent distributions of preference to an aggregate distribution of preferences has an obvious benefit: reducing the dimensionality of the numerical integration during model estimation. Since MAPL models integrate over the univariate distribution of utilities rather than the multivariate distribution of parameters as in mixed logit models to determine choice probabilities, the simulation load from numerical integration can be significantly reduced. Furthermore, a MAPL model does not limit modelers to using convenient mixing distributions that are often used in mixed logit. As we illustrate in the simulation experiment in Section \ref{sec:SimulationExperiment}, a MAPL model is capable of achieving greater flexibility in modeling unobserved heterogeneity by using alternative mixing distributions, such as the semi-parametric distribution posed by \citet{Fosgerau2013}.

\newpage

\noindent
\textbf{Declaration of generative AI and AI-assisted technologies in the writing process}

\noindent
During the preparation of this work the author(s) used the Claude 3.5 Sonnet Large Language Model by Anthropic to improve the clarity of language in some paragraphs. After using this tool/service, the author(s) reviewed and edited the content as needed and take full responsibility for the content of the publication.

\newpage

\end{document}